\documentclass[12pt, multphys,vecphys]{svmult}

\usepackage{amsfonts}        % American Mathematical Society fonts
\usepackage{amssymb}
\usepackage{natbib}
\usepackage{cite}
\usepackage{makeidx}         % allows index generation
\usepackage[bottom]{footmisc}% places footnotes at page bottom

\makeindex             % used for the subject index
\begin{document}

%  TITLE

\title*{Cold Feedback in Cooling--Flow Galaxy Clusters}

\titlerunning{Cold Feedback}

% AUTHOR

\author{Fabio Pizzolato}

\authorrunning{Pizzolato}

\institute{
Department of Physics\\ 
Technion--Israel Institute of Technology\\
Haifa 32000\\
\texttt{fabio@physics.technion.ac.il}
}

\maketitle

% ABSTRACT

\begin{abstract}
We put forward an alternative view to the Bondi--driven feedback 
between heating and cooling of the intra-cluster medium (ICM) 
in cooling flow (CF) galaxies and clusters.

We adopt the popular view that the heating is due to an
active galactic nucleus (AGN), i.e.  a central black hole
accreting mass and launching jets and/or winds.
We propose that the feedback occurs with the {\em entire}
cool inner region ($r \lesssim 5-30 {~\:\rm kpc}$).  A  moderate
cooling flow {\em does} exist here, and  
non--linear over--dense blobs of gas cool fast and are removed
from the ICM before experiencing the next major AGN heating event.
Some of these blobs may not accrete on the central black hole, but may  
form stars and cold molecular clouds.
We discuss the conditions under which  the dense blobs 
may cool to low temperatures and feed the black hole.

\end{abstract}

%%%%%%%%%%%%%%%%%%%%%%%%%%%%%%%%%%%%%%%%%%%%%%%%%%% MAIN BODY

\section{Introduction}
\label{s:intro}

The observations of galaxy clusters with the last generation
X-ray satellites {\sl Chandra} and {\sl XMM-Newton} failed to 
detect  the large amounts of cool gas predicted by
the old version \citep{Fab94}  of the  cooling flow (CF)
model  (see e.g. \citealp{Pet06} for a recent review).
The most straightforward explanation is that the intra-cluster medium (ICM)
in CF clusters is  heated by some mechanism, the currently most 
popular candidate being  the  active galactic nucleus
(AGN) residing at the core of the cluster dominant  galaxy.

Most models of AGN heating agree in that there is some 
feedback between the heating and the radiative cooling, 
possibly resulting  in intermittent AGN activity.
There are two approaches to the AGN/ICM feedback.
In the {\it hot feedback} scenario
(see e.g. \citealp{Nul04})  the ICM never  cools
below X--ray emitting temperatures; the AGN
accretion pattern is Bondi--like, and is determined by the ICM properties
at the Bondi radius, which is typically few tens ${~\:\rm pc}$.

In the second approach, dubbed {\it  cold feedback}, 
(see \citealp{Piz05} and \citealp{Sok06} for more details),  
the black hole accretes {\em cold}  gas, although 
the amount of cooling mass  is much below that predicted by the old
cooling flow model.
In this case the feedback takes place within  a region
extending to a distance of $\approx 5-30{~\:\rm kpc}$ from the cluster centre.

\section{The Cold Feedback Scenario}
\label{s:scenario}

The cold feedback scenario entails a cycle in the cooling/accretion
activity.

We suggest that this cycle starts with a major AGN outburst,
which injects a huge amount of energy into the ICM.
The AGN outburst interacts in a very complicated fashion with the ICM
\citep{Beg04}, e.g. it heats and inflates radio bubbles, and
may also stir some turbulence.
The ICM itself is displaced  and thickened by the rising bubbles,
as shown by their rims'  enhanced X--ray brightness \citep{Bla01}.
A non-homogeneous thickening  may result in the formation of a 
multi-phase gas, consisting e.g. of pockets of cold gas  with a wide 
spectrum of densities. Owing to their over-density, these blobs
fall to the black hole.

If these these blobs have a significant amount of  
angular momentum, they cannot  feed the black hole.
On the other hand, low angular momentum blobs 
are allowed to accrete, igniting a fresh AGN outburst, which in
turn disturbs the ICM, restarting the cycle with a new
injection of  blobs.

In this duty cycle 
some of the gas cools to low temperatures ($\lesssim 10^4 {\rm K}$) 
before the next major heating episode, while the rest is heated back to
a relatively high temperature.
The presence of a detectable amount of gas cooling  below X-ray
emitting temperatures  is a specific prediction of this model.
Indeed, in the CF cluster A~2597 \index{A2597} both extreme-UV and X-ray
observations indicate a mass cooling rate of $\sim 100 {~\:M_\odot\;\rm yr^{-1}}$,
which is $\sim 0.2$ of the value quoted in the past based on {\sl ROSAT} 
X-ray observations (see the discussion in \citealp{Mor05}).
In the CF cluster A~2029\index{A2029}, \citet{Cla04} find a substantial
amount of gas at a temperature of $\approx 10^6 {\rm K}$; a CF model gives
a mass cooling rate of $\sim 50 {~\:M_\odot\;\rm yr^{-1}}$.

According to  the results of \citet{Piz05} and \citet{Sok06}, a key role in 
the cold bubbles' accretion  {\em and feedback} 
is played by the ICM entropy profile.

Long after an outburst, following an extended period of cooling, the
entropy profile is  steep. In this case it is difficult for an infalling  
blob  to accrete and feed the black hole: on its way it will
reach an equilibrium radius where
its density equals that of its surroundings;  it  then
dissolves   {\em before} accreting on the black hole. 
Only the blobs the birthplace of which has a small entropy difference 
with respect to the core can accrete. Therefore, the  first accretion 
episodes are most likely to involve small blobs, stemming not far from 
the centre. The AGN activity induced by their accretion is rather weak,
and only raises the entropy very close to the AGN. In the meantime,
the ICM further out keeps cooling, reducing the cooling time of
the dense blobs. In addition,  the combined action of  cooling 
the far regions and  heating the central one flattens the cluster's
entropy profile. By the same  token, due to the flat
entropy profile now  even far blobs may
accrete on the AGN, triggering a major outburst. This may therefore heat
the cluster on large scales.

\bigskip

There are  four  hurdles  the blobs must  overcome  before accreting.
\begin{enumerate}
\item
Some pockets may be engulfed by the expanding radio lobes. 
As \citet{Piz05} (and references therein)
argue, if the blobs are initially dense enough 
(say, $10-100$~times the ambient medium) they can survive the shock, and
accrete unhindered through the radio lobes.
\item
These blobs must withstand the ICM thermal conduction: if it is
too efficient, it would be able to  evaporate these 
cold clouds  {\em before} they can accrete on the AGN.
We find from Figure~3 of  \citet{Nip04} that the effective heat conduction 
should be $\lesssim 10^{-3}$ times the nominal \citet{Spi56} not 
to  evaporate a blob of radius  $a \sim 10-100 {~\:\rm pc}$.
Such a strong suppression factor is supported by theoretical 
considerations \citep{Pis96, Nat03}. The possible existence of
magnetic turbulence \citep{Sch06}   may also strongly affect 
the transport coefficients, including thermal conductivity.
A somewhat suppressed conduction  is also consistent with some
recent observations (M87\index{M87}: \citealp{Mol02}, NGC~5044: \citealp{Buo03}).
\item
As \citet{Nul86} pointed out,
{\it in the absence of a cohesive force}
a blob would be torn apart  by  the ram pressure
in~$\sim 10^7{~\:\rm yr}$, i.e. a time considerably shorter
than the time taken by the blob to fall to the centre. Some kind
of cohesive force (like a magnetic tension) must then be  at work
to prevent the blob disruption. 
\item
If the blobs' angular momentum is too high, they are prevented from
approaching the central black hole: the flux would merely stagnate,
cool down and condense in filaments or stars, and the 
AGN fuelling is cut off altogether, thus  making  the feedback impossible.
Indeed,  the existence of a circumnuclear disc with radius 
$R_d \approx 10^2{~\:\rm pc}$ around M87 
\citep{Har94, For94} shows that
the flow possesses an amount of angular momentum.

However, the blobs are expected to 
form and accrete only in a region of the same extension as the inner
gas entropy  plateau ($\sim 5-30{~\:\rm kpc}$). 
The circularisation radius of this flow is expected to be 
of the same order as the  actual size of the circumnuclear
disc of M87 \citep{Piz05}, i.e. direct accretion is possible  down to  
the immediate  vicinity of the black hole.
In addition, the blobs  may stem  directly from  ICM
disturbances  driven by an early AGN activity,
but also  from galaxies mass-stripping \citep[][]{Sok91}.
Since the  galaxies do not have an ordered bulk motion, also the blobs
stripped from them are  also unlikely to  organise in an
ordered  flow with high net angular momentum. 

\end{enumerate}

So, to summarise, under some reasonable assumptions the cold blobs are
able to survive long enough to accrete on  the AGN and hence provide feedback.

The dense blobs that sink to the centre feed the AGN.
The feedback is with the {\it  entire} cool inner region, and not
only with the gas close to the black hole.
Any over-cooling taking place  in the inner region, where the
temperature profile is flat, will lead to many small and dense blobs, which
feed the AGN.

A Bondi accretion radius  as large   as  the
disc around the black hole of M87  further suggests that
the simple Bondi accretion flow
\citep[][]{Chu02, Nul04} does not hold; the accreted material
has a larger angular momentum, and may come from much larger radii.
Also for cluster A~1835 \index{A1835} the Bondi 
accretion is unlikely to be the 
main engine powering the feedback \citep{McN06}.

\section{Summary}
\label{s:summary}

We propose that the feedback occurs with
the entire cool inner region, $r \lesssim 5-30 {~\:\rm kpc}$, in
what we term a {\em  cold-feedback model}.
In the proposed scenario non-linear over-dense
blobs of gas cool fast and  are removed from
the ICM before  the next major AGN heating event in their region.
It is important to note that an AGN burst can take place and
heat other regions, since the jets and/or bubbles may
expand in other directions as well.
The typical interval between such heating events at a specific region
is~$\approx 10^8 {~\:\rm yr}$.
Some of these blobs cool and sink toward the central black hole,
while others may form stars and cold molecular clouds.

Four conditions should be met in the inner region participating
in the feedback heating.

\begin{enumerate}

\item 
In order for the blob not to reach an equilibrium point before
accreting,  a shallow  ICM entropy profile is required.
The relevant dense blobs then must form  within the cluster core,
typically  $\lesssim 5-30{~\:\rm kpc}$ from the centre.
We note that the lower segment of magnetic flux loops can
be prevented from reaching the stabilising point by the upward
force of the magnetic tension inside the loop \citep[][]{Sok04}.
Therefore, some perturbations can be formed at large distances,
where density profile is steep, and still cool to low temperature
and feed the central black hole.

\item 
Non-linear perturbations  are required. These presumably formed mainly
by previous AGN activity, e.g.  jets and radio lobes.

\item 
The cooling rate  of these non-linear perturbations
is short relative to few times the typical interval  between
successive AGN outbursts.

\item
The blobs must not be evaporated by thermal conduction
before they are delivered to the AGN. This requires a strong
suppression of thermal conduction.

\end{enumerate}

The first and the third condition, which are not completely
independent of each other,
require that the initial ICM cools by a factor of a few before the
feedback starts operating, and the second condition requires that the
inner region must be disturbed.

The cold-feedback model has the following implications and predictions
\begin{enumerate}

\item
The optical filaments
observed in many CF-clusters and the cooler molecular
gas detected via CO observations come from cooling ICM
(with some amount possibly from stripping from galaxies).

\item
Some X-ray emission from gas at
temperatures $\lesssim 10^7{\rm K}$ is expected, consistent with the
moderate CF model.  This is
much more than in many other AGN heating models, 
but at least  an order of magnitude below what predicted by
the old CF model,
We stress that in the cold-feedback heating, cooling flows
{\it  do exist}. 
Such gas cooling to below X-ray emitting temperatures was
found recently in two CF clusters 
(A2597\index{A2597}: \citealt{Mor05}; A2029\index{A2029}: \citealt{Cla04}).

\item
The feeding of the central black hole with cold gas in the
cold feedback models makes the process similar in some
aspects to that of AGN in spiral galaxies.
Therefore, the outflow can be similar \citep[][]{Sok05}.

\item
It is possible that in the cold feedback model a
substantial fraction of gas that cooled to low
temperatures and was accreted to the accretion disc around
the central black hole, is injected back to the ICM at
non-relativistic velocities \citep{Sok05}.

\end{enumerate}

% REFERENCES

%\nocite{}
%\bibliographystyle{apj}
%\bibliography{referenc}

\printindex

\end{document}